# Policy-Aware Mobility Model Explains the Growth of COVID-19 in Cities

Zhenyu Han[1&], Fengli Xu[34&], Yong Li[1*], Tao Jiang[2], Depeng Jin[1], Jianhua Lu[1], James A. Evans[345*]

1 Beijing National Research Center for Information Science and Technology (BNRist), Department of Electronic Engineering, Tsinghua University, Beijing, P. R. China
2 School of Electronics Information and Communications, Huazhong University of Science and Technology, Wuhan, P. R. China
3 Knowledge Lab, Department of Sociology, University of Chicago, Chicago, IL, USA
4 Knowledge Lab, University of Chicago, Chicago, IL, USA
5 Santa Fe Institute, Santa Fe, NM, USA
& Equal Contribution; * e-mails: liyong07@tsinghua.edu.cn, jevans@uchicago.edu

**With the continued spread of coronavirus, the task of forecasting distinctive COVID-19 growth curves in different cities, which remain inadequately explained by standard epidemiological models, is critical for medical supply and treatment[1,2]. Predictions must take into account non-pharmaceutical interventions to slow the spread of coronavirus, including stay-at-home orders, social distancing, quarantine and compulsory mask-wearing, leading to reductions in intra-city mobility and viral transmission[3]. Moreover, recent work associating coronavirus with human mobility[4,5] and detailed movement data[6,7] suggest the need to consider urban mobility in disease forecasts[8,9]. Here we show that by incorporating intra-city mobility and policy adoption into a novel metapopulation SEIR model, we can accurately predict complex COVID-19 growth patterns in U.S. cities ($R^2 = 0.990$).**

**Estimated mobility change due to policy interventions is consistent with empirical observation from Apple Mobility Trends Reports[10] (Pearson's R = 0.872), suggesting the utility of model-based predictions where data are limited. Our model also reproduces urban "superspreading"[11], where a few neighborhoods account for most secondary infections across urban space, arising from uneven neighborhood populations and heightened intra-city churn in popular neighborhoods. Therefore, our model can facilitate location-aware mobility reduction policy that more effectively mitigates disease transmission at similar social cost. Finally, we demonstrate our model can serve as a fine-grained analytic and simulation framework that informs the design of rational non-pharmaceutical interventions policies.**

In response to the rapid development of the COVID-19 pandemic and the lack of a widely available vaccine, local health authorities worldwide have considered, debated, and implemented a wide range of non-pharmaceutical intervention policies that are deemed both necessary and effective to contain the spread of coronavirus[12,13]. Human mobility control has become the centerpiece of the most common intervention measures, including cancellation of social gatherings[14–16], quarantine and stay-at-home orders[17], and suspension of public transportation[18]. Nevertheless, complex and distinctive growth patterns associated with COVID-19 cases have ranged from exponential to sublinear growth across cities[19–21], requiring epidemiological models to be expressive enough to account for different policy adoptions with their efficacy. Moreover, recent studies have found accumulating evidence of "superspreading events" that suggest the spatial heterogeneity of coronavirus risk in urban space and contradict the assumption of a homogeneously mixing population underlying classic epidemiological models such as SEIR[22]. These developments pose the pressing need for a simple while general epidemiological model that can capture heterogeneous urban mobility and the varying effect of public health intervention policies across cities to accurately predict COVID-19 cases growth patterns even when detailed infection distribution and mobility data are unavailable.

Here, we present a metapopulation SEIR model that explicitly incorporates intra-city mobility patterns and systemic policy adoption. We model the urban space as a set of spatially distributed neighborhoods, and use a general gravity model[23] to capture intra-city mobility behavior as mobility flows among neighborhoods. Inspired by Stouffer's law of opportunity-driven urban movement[24], our gravity model predicts mobility flow as proportional to neighborhood population and inversely proportional to the travel distance. Atop the predicted mobility flows, we overlay a metapopulation SEIR model to characterize the infection within each neighborhood and track inter-neighborhood disease spread arising from mobility. This model allows us to jointly consider different urban mobility patterns and local intervention policies in a principled way that facilitates the prediction of complex COVID-19 growth patterns. Recent works demonstrated the power of mobility data, when available, to forecast COVID-19 incidences[5–7,25]. We extend this line of research by considering a simple yet powerful urban mobility model that can be easily generalized to any country or region *without* empirical intra-city mobility data.

Simulation of the 20 most infectious U.S. counties demonstrates that our model can accurately predict complex and distinct growth curves tracing COVID-19 cases in cities ($R^2$ =0.990). Our estimates of mobility change due to non-pharmaceutical interventions are consistent with real-world observation derived from Apple Mobility Trends Reports[10] (Pearson's R = 0.872), suggesting the power of our model-based predictions even when human mobility data are lagging or limited. Our model can also characterize the "superspreading" events in urban space through tracing the coronavirus spread among urban neighborhoods, where a small portion (20%) of neighborhoods account for a large portion (68.3%) of secondary infections. Moreover, model analysis reveals that urban "superspreading" is a joint result of uneven urban population and heightened intra-city churn associated with popular neighborhoods. We demonstrate that by focusing on the regions predicted to have the highest infection risk, location-aware mobility reduction policy can result in significantly more effective epidemic control at similar social cost. This suggests the possibility of flexible and targeted urban epidemic control policies. Moreover, our model can serve as a simulation framework for the cost-effective evaluation of various hypothetical intervention measures. Simulation results show the widely adopted mobility reduction, social distancing and quarantine policy have significant effect on curb the transmission of coronavirus, and it is important to put forward timely responses with adequate implementation. Overall, our model facilitates the fine-grained analysis, evaluation, and rational design of non-pharmaceutical interventions policies for urban space.

**Policy-aware metapopulation model for COVID-19 growth curve prediction**

We use a metapopulation scheme to model the fine-grained disease spreading dynamics within cities. This breaks down urban space into numerous neighborhoods with a certain population, and maintains a separate SEIR model with its own susceptible (S), exposed (E), infectious (I), and recovered (R) states for the subpopulation within each neighborhood (see Extended Data Fig. 1). We use a gravity model to characterize mobility flows between neighborhoods, mainly considering the cost of travel and the attraction of social opportunity[22](see Methods M1), i.e., the mobility flows are predicted to be proportional to neighborhood population and inversely proportional to the travel distance. Therefore, secondary infections occur within each neighborhood based on simulation of the SEIR model, and disperse to other neighborhoods according to the predicted mobility flows. We set three free parameters to capture the efficacy of popular non-pharmaceutical intervention policies: (1) a learnable infection rate $\beta$ to account for social distancing policies, e.g., limiting social gathering[12,13]; (2) a learnable quarantine rate $\kappa$ to account for the capacity of testing and quarantine[17,26]; (3) and a learnable mobility level $M$ to account for mobility reduction policies, e.g., stay-at-home orders and suspension of public transportation[18]. Other epidemiological parameters are set according to recent COVID-19 studies (see Extended Data Table 1). Urban population distribution is derived from the open source WorldPop database[27]. To characterize the behavioral change due to intervention policies, we segment the simulation period into two parts based on the announcement of nationwide emergency status in the U.S. and fit separate parameters for each period (Extended Data Fig.2). We calibrate the parameters of our model based on official counts of confirmed cases[28].

We evaluate our model with the task of reproducing growth curves of COVID-19 confirmed cases in 20 U.S. counties from the beginning of the pandemic through April 30. These counties, often coterminous with U.S. cities (e.g., Chicago and Cook County), demonstrate complex and distinctive growth patterns (Fig.1a, red dot lines), which can be categorized into four types of increase: linear (Hudson, King, etc.), concave (Bergen,

Miami-Dade, Nassau, etc.), convex (Cook, Davidson, Los Angeles, etc.), and S-shaped—convex then concave (Harris, New Orleans, Will). All these distinct curves can be accurately reproduced by our model, with $R^2$ scores above 0.99 in all cities (Fig.1a, blue lines). On the contrary, the standard SEIR model fails to reproduce these empirical growth curves, only predicting exponential or no growths with $R_0 > 1$ and $R_0 < 1$, respectively (Fig.1a, green lines)[1]. Moreover, we theoretically prove that our model is able to reproduce complex growth curves with different intra-city mobility levels (see Methods M2), which greatly extend the expressive power of classic epidemiological models. Further results echo this finding that different levels of intra-city mobility have a significant impact on the shape of city-level growth curves (Extended Data Fig.3). These results justify the need to incorporate intra-city mobility behavior to characterize complex and distinctive growth curves of COVID-19 cases, and provide a microscopic foundation to rationalize them from the perspective of human behavior. We further evaluate our model by predicting future confirmed cases within two weeks, i.e., May 1-14 (Fig.1b). The results demonstrate that our model significantly outperforms the standard SEIR model with reduced normalized root mean square error (NRMSE) from 7.222 to 0.294. When we perform a correlation analysis on the mobility reduction estimated by our model and real-world observations derived from Apple Mobility Trends Reports[10] (Fig.1c), our model accurately estimates behavior changes by solely observing the growth curves of confirmed cases (Pearson's R= 0.872). These results suggest that our model can be robustly generalized to cities and regions where mobility data is lagging or limited.

**Reproducing and rationalizing the superspreading events in urban space**
The events of superspreading have been widely observed in epidemics like SARS, Measles and Smallpox[22], as well as COVID-19, where a small portion of infected people and locations are responsible for a disproportionate number of secondary infections. Researchers have identified accumulating superspreading events of COVID-19 through case study[29,30], phylogenetic analysis[17,31] and statistical evaluation[32,33]. Despite its importance, superspreading cannot be adequately explained by the standard SEIR model, because of the fundamental assumption of homogeneous population mixing. We demonstrate that the population distribution in a city mostly follows an exponential pattern with high Gini index (Extended Data Fig. 4). As a result, researchers proposed to characterize the superspreading events with dispersion parameter $k$, which measures how the transmissive power of each individual deviates from the general population[22,30] with a negative binomial model. However, this statistical point of view reveals limited understanding about the underlying behavioral mechanism of superspreading, e.g., intra-city mobility behavior. Moreover, recent research found evidence that superspreading events can be better characterized with fat-tailed power law distribution[33], which further suggested its link to human behavior patterns. To bridge this gap, we examine the mechanism of superspreading events through the analytic framework provided by our model, tuned to analyze the spatial heterogeneity of intra-city disease spreading dynamics.

To evaluate the infection risk in different neighborhoods, we use the infectee-infector ratio as a proxy to examine the selected U.S. counties, which is defined as the average number of secondary infections per infected person in each neighborhood. We rank the neighborhoods based on the infectee-infector ratio, and calculate the cumulative distribution function of the number of infected and secondary infections (Fig. 2a). If the infection risk is spatially homogeneous in urban space, we expect the cumulative distribution function grows linearly as the black dot line. However, our model reproduces a highly skewed distribution where the most infectious 20% of neighborhoods are responsible for 68.3% of secondary infections. The Gini index also captures the unevenness in the spatial distribution of secondary infections and infected people[34], with a score of 0.630 and 0.663, respectively. More importantly, the superspreading events reproduced by our model have similar statistical distribution with real-world observations, where the simulated dispersion parameters are 0.208~0.215 (see Extended Data Fig.5) falling in the empirically observed range of 0.147~0.667[35]. It indicates our model can effectively extend the expressive power of the standard SEIR model in terms of adequately reproducing the empirically observed superspreading events[35].

Recent research pointed out that urban population distribution[36] and intra-city mobility behavior[25] might be related to the superspreading events. Thus, by leveraging our model as a simulation framework, we aim to shed light on how these two factors impact on the superspreading. The effect of uneven urban population distribution is examined in Fig.2b and c, where the intra-city mobility factor is controlled by replacing the gravity model with random movement. We can observe that if the population is distributed evenly in urban space, then the superspreading events will completely disappear (Fig.2b). On the contrary, if we preserve the real-world population distribution, the superspreading events can be reproduced, but of a much lower level with the Gini index of 0.198 and 0.090 respectively (Fig.2c). These observations combined to suggest that the uneven distribution of urban population is indeed an important factor, but it only accounts for a portion of the overall superspreading events, which leads us to examine the role of intra-city mobility behavior. The gravity model we use to capture the intra-city mobility mainly considers two factors, *i.e.*, the decreasing travel probability with distance and the increasing travel probability with population. To examine the role of each factor, we evaluate two variants of mobility model that only keep one factor respectively. When we combine real-world population distribution and the mobility model that only considers the distance factor, our model reproduces a similar level of superspreading events as the combination of real-world population distribution and random movement (see Fig.2d). Besides, if we adopt the mobility model that only considers the population factor, we reproduce a strong superspreading effect with Gini index of 0.632 and 0.539 for secondary infections and infected persons (see Fig.2e), which are similar with the results of the complete model in Fig.2a. Therefore, these results suggest that the superspreading events are a joint result of uneven urban population distribution and heightened intra-city churn associated with popular neighborhoods.

**Evaluating location-aware mobility reduction policy in U.S. counties**
Mobility reduction policy has been one of the most adopted non-pharmaceutical interventions that aims to curb the coronavirus spread in urban space[7,37]. However, large-scale mobility reduction schemes like city-wide lockdowns often cause great socioeconomic consequence[38]. Since there is high spatial heterogeneity of infection risk in urban space, a promising solution is to design location-aware, targeted mobility reduction policy that achieves better trade-off between epidemic control and social cost. Here, we investigate this possibility through the simulation of limiting the intra-city mobility of several selected neighborhoods. We control the social cost as the overall population size of selected neighborhoods (5% of the city population), and evaluate the policy efficacy as the percentage of confirmed cases that are prevented.

Highly populated neighborhoods are inherently more vulnerable to coronavirus transmission[36]. Thus, a naive solution will be to control the mobility of the top populated neighborhoods. However, due to intra-city mobility behavior, the infection risk does not solely depend on the population size. Although the population size and infectee-infector ratio of neighborhoods have a relatively high correlation (spearman correlation = 0.74), the infectee-infector ratio still has a significant portion of variance that cannot be adequately explained by population size (Fig.3a). Specifically, if we choose to control the mobility of the top-populated neighborhoods, 21.0% of overall confirmed cases can be avoided, which outperforms the random selection baseline (Fig.3b). On the contrary, if we select the most infectious neighborhoods predicted by our model, the overall infection can be significantly reduced by 44.8% (Fig.3b). These results suggest our model can indeed inform the design of more cost-effective location-aware mobility reduction policy.

To examine the source of different efficacy, we visualize the selected neighborhoods in two counties as case study (Fig.3c and d). The case of Los Angeles county (L.A.) shows the baseline policy focused only on population density would miss the high infection risk that occurs in less populous neighborhoods around central L.A., because it cannot model these neighborhoods' mobility flows with the nearby and densely populated neighborhoods (Fig.3c). Similarly, our model predicts that the peninsulas surrounding the Tampa area has a smaller infection risk since it has longer distance with most populated neighborhoods (Fig.3d). More detailed analysis suggests the location-aware mobility reduction policy informed by our model can also better alleviate the superspreading events (Extended Data Fig. 6).

**Quantifying the effect of intervention policies**
To estimate the effect of various intervention policies on controlling coronavirus spread,

we evaluate our model in 20 U.S. counties with different parameter settings that simulate the implementation of the widely adopted interventions, e.g., mobility reduction, social distancing, and quarantine (Fig.4a-c). Simulation shows 30% higher intra-city mobility will lead to nearly threefold overall infections (2.93 times; IQR 1.57-7.24), and 30% less intra-city mobility will halve the number of cases (0.525 times; IQR 0.237-0.783). The average observed mobility drop in U.S. counties during the pandemic is around 30.3% (Extended Data Fig.7, with IQR 0.227-0.392), which suggests that current mobility behavior changes have been effective in reducing coronavirus transmission. Besides, a 10% infection rate increase multiplies citywide infections by 2.10 (interquartile range, IQR 1.51-3.30), while a 10% decrease almost halves the infections (0.524 times; IQR 0.231-0.661). This suggests the great importance of reducing the infection rate of coronavirus, which can be achieved through the implementation of personal hygiene and social distancing policies, such as wearing masks and cancelling large gatherings[39,40]. Quarantine is another widely adopted policy to contain the secondary infection of confirmed and suspected cases[41]. Our model predicts that the number of confirmed cases will be 29% higher given a 10% decrease in quarantine rate (1.29 times; IQR 1.06-2.38), while a 10% increase in quarantine rate will prevent approximately 78% of citywide infections (0.221 times; IQR 0.0739-0.525). This asymmetric effect of quarantine rate suggests the necessity of ensuring that testing capacity and hospital resources are sufficient. These capacities can greatly reduce overall infections once they pass a critical threshold. Finally, timely responses to COVID-19 growth are often considered critical in combating the virus[42]. To explore this effect with our model, we shifted the time point of behavior change to simulate the potential effect of different response time. We find that a delay of 10 days causes 3.26 times higher overall infections (IQR 1.58-9.53), while 10 days earlier reduces the infections by 57.5% (0.425 times; IQR 0.12-0.71). These experiment results indicate the widely adopted interventions, like mobility reduction, social distancing and quarantine, and timely responses play important roles in curbing virus spread and flattening the growth curves.

**Discussion**

Our model aims to reconstruct the fine-grained transmission process of COVID-19 in cities with minimum real-world data. We show that by only observing the population distribution and the number of confirmed cases, our model can accurately reproduce the complex growth curves of COVID-19 cases and forecast the future trends. Our model provides a theoretical framework to explain the distinctive growth curves in different U.S. urban counties from the perspective of intra-city mobility. Furthermore, our model can reproduce and explain the superspreading events in cities resulting from the uneven distribution of urban population and the assumption of spatially heterogeneous mobility. In these ways, our model markedly augments the expressive power of current epidemiological models. Besides, the minimum data requirement ensures it can be easily generalized to cities and regions without fine-grained mobility data.

There are two limitations of our study. First, our parsimonious model only incorporates three learnable parameters that account for infection rate, quarantine rate and mobility level. We do not consider all fine-grained features that might influence the spread of coronavirus, such as weather[43], changing attitudes towards containment policies[44], and population demographics like age[26] and socioeconomic status[6]. Second, the embedded mobility model reconstructs intra-city mobility in aggregate as the mobility flux between neighborhoods, abstracting over individuals and unable to pinpoint specific infections. Nevertheless, analysis demonstrates that our model can accurately trace citywide COVID-19 cases, suggesting its expressive sufficiency for characterizing urban viral transmission dynamics. Moreover, the minimum model setting improves its robustness and generalizability.

Our results have several implications for non-pharmaceutical and public health interventions and benefit the design of reopening policies. Our analysis suggests that the widely observed superspreading events are not solely due to the randomness of rare events, but also systematically linked with uneven urban population distribution and the disproportional mobility attractiveness of populous neighborhoods. It further demonstrates the need to jointly consider urban environment and human mobility behavior to effectively curb superspreading events. We also show how infection risk is spatially heterogeneous in urban space, and our model can predict the most infectious locations. This suggests the possibility of designing targeted and location-aware mobility reduction policies that achieve better performance with identical social cost. Next, model analysis demonstrates that estimated mobility change is consistent with real-world observations in Apple Mobility Trends Reports. This affirms that our model can offer insights into fine-grained urban transmission dynamics and be robustly transferred to cities without mobility data. Finally, our model can be used to comparatively and inexpensively evaluate the effect of public health interventions, including social distancing, stay-at-home orders, quarantine, and dynamically customized mobility reduction policies that substantially curb the spread of coronavirus.

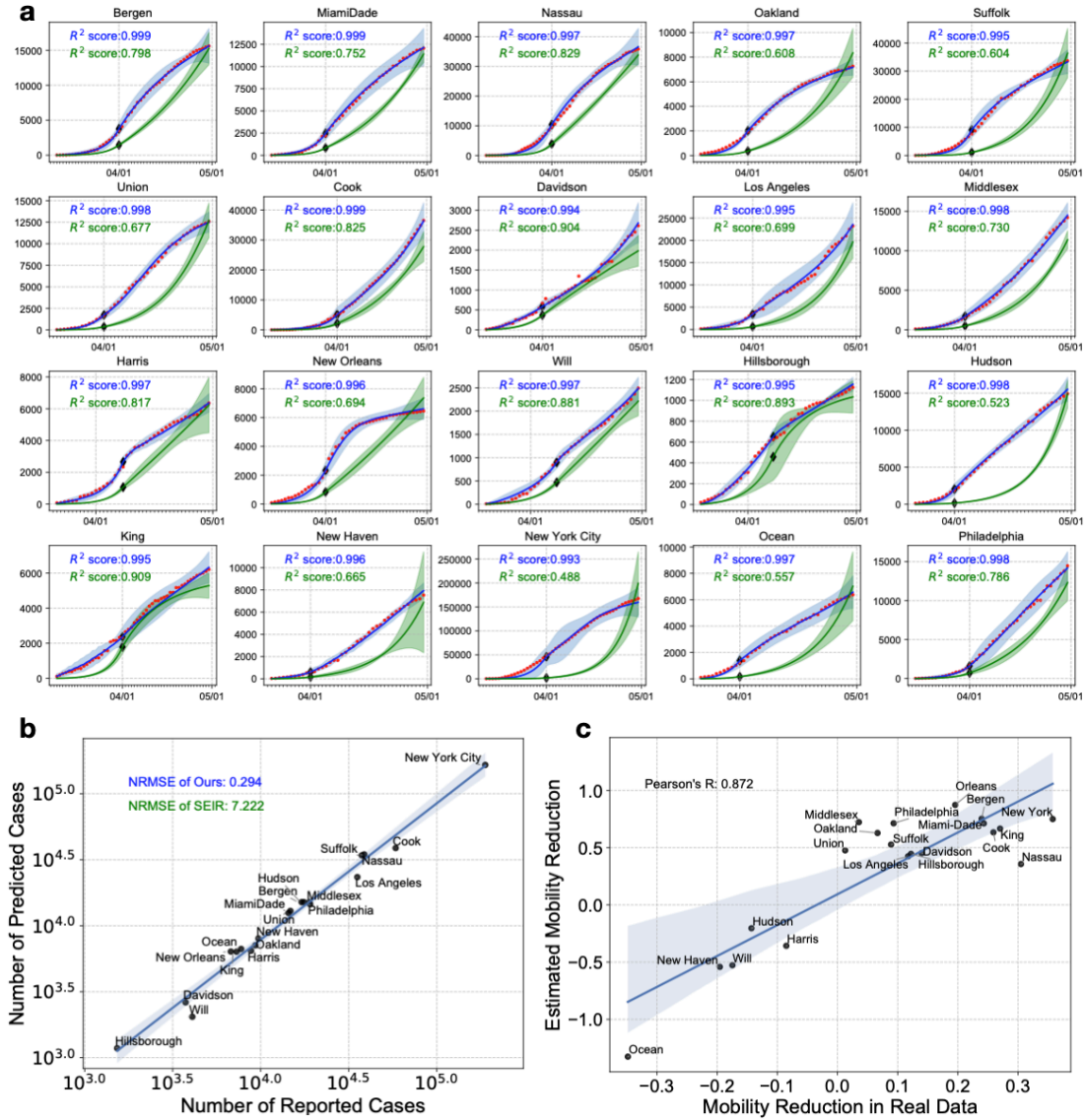

**Fig.1| Predicting COVID-19 dynamics in the 20 most infectious U.S. urban counties. (a)** Reproducing the growth curves of COVID-19 confirmed cases. Red dots are empirical observations of confirmed cases, blue lines are growth curves estimated by our model, and green lines are growth curves estimated by the standard SEIR model. The shaded area represents the 99% confidence interval. Diamond markers denote the time point of behavior change. Our model can better reproduce empirically observed COVID-19 growth curves, results in a significantly higher $R^2$ score. **(b)** Evaluating our model with the task of predicting future confirmed cases within 14 days. The normalized root mean square error (NRMSE) of our model is 0.294, compared to 7.222 for the standard SEIR model. **(c)** Correlation between empirical mobility reduction and model estimation before and after announcement of nationwide emergency status around March 25. Our model can accurately estimate the behavior change with Pearson correlation coefficient as 0.872.

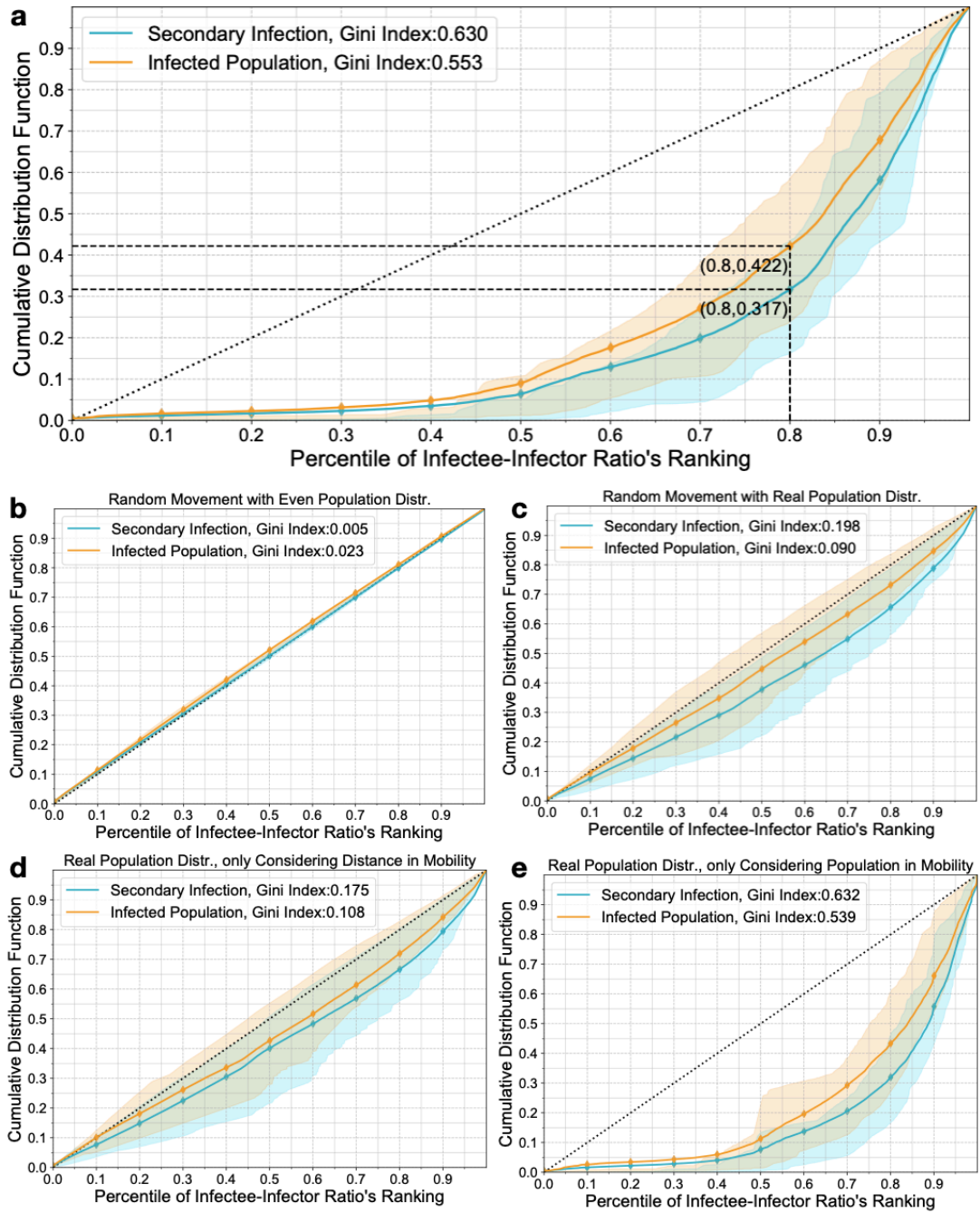

**Fig.2| Reproducing and rationalizing the superspreading events in urban space. (a)** The distribution of secondary infection and infected persons across neighborhoods ranked by infectee-infector ratio. If the secondary infection and infected persons are distributed evenly, they are expected to follow the $y = x$ curve (dotted line). Our model reproduces a significant uneven distribution with the bottom 80% neighborhoods only accounting for 42.2% infected persons (orange line) and 31.7% secondary infections (blue line). Shaded areas represent interquartile range for all urban counties. The Gini index for the distributions of secondary infections and infected persons are 0.630 and 0.553, respectively, which suggests the infection risk is spatially

heterogeneous in urban space. **(b)** The uneven distribution of secondary infections and infected persons disappears when we replace the urban population and mobility model with even distribution and random movement, respectively. **(c)** The results with moderate level of uneven distribution when combining real-world population distribution with random movement. **(d)** When combining real-world population distribution with the mobility model that only considers the distance factor, we reproduce a similar level of uneven distribution as (c). **(e)** When combining real-world population distribution with the mobility model that only considers the population factor, we reproduce a similar level of uneven distribution as the complete model in (a).

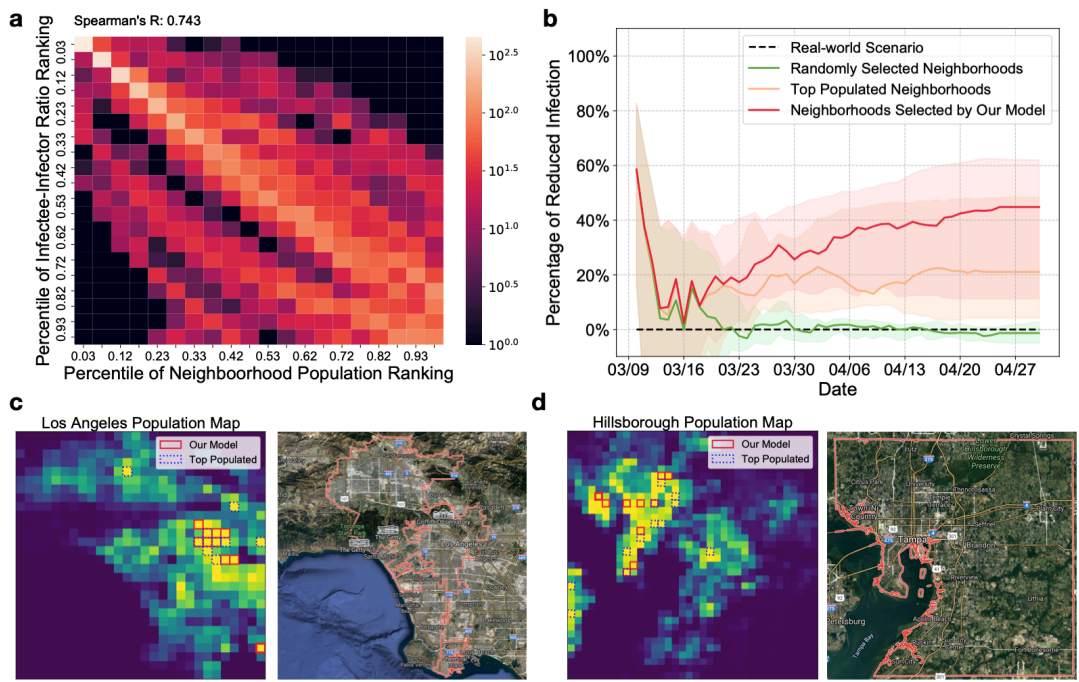

**Fig.3| The effectiveness of different mobility reduction policies. (a)** The heatmap of neighborhood's ranking on population size and infectee-infector ratio across 20 counties, which has a relative high spearman correlation of 0.743, but also a high variance in infectee-infector ratio that cannot be explained by population size. **(b)** Performance of different mobility reduction policies. The policy informed by our model leads to 44.8% decrease in cumulative number of infections, which significantly outperforms the strategy of selecting top populated neighborhoods and random selection (the solid lines denote the median value of all counties, and shaded areas represent interquartile ranges). **(c)** Visualization of selected neighborhoods in Los Angeles county. **(d)** Visualization of selected neighborhoods in Hillsborough county.

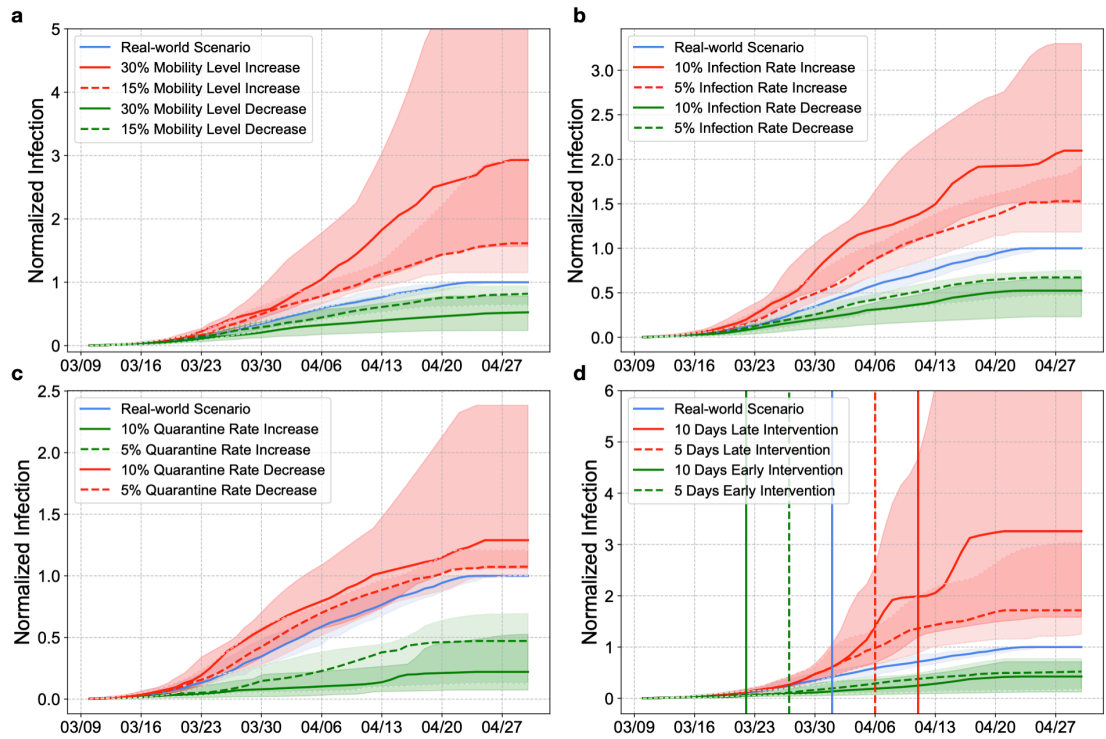

**Fig.4| Estimated growth curves of COVID-19 infections in U.S. counties under different intervention scenarios.** All Figures show the median value of confirmed cases across 20 U.S. counties under different scenarios normalized by the overall real-world confirmed cases, where the shaded areas denote the interquartile range. **(a)** The estimated growth curves with different intra-city mobility levels. **(b)** The estimated growth curves under different quarantine rates. **(c)** The estimated growth curves under different infection rates. **(d)** The estimated growth curves with different intervention dates. The vertical lines denote the dates of intervention policy.

**Methods:**

**M1 Meta-population SEIR model**

The standard SEIR models use ordinary differential equations (ODE) to trace the epidemic development, which divides the population into four status: susceptible (S), exposed (E), infected (I) and recovered (R). Besides, one fundamental assumption behind the standard SEIR model is homogeneous population mixing, and hence each susceptible individual will have similar infection risk. As a result, it can neither reproduce the complex growth curves of COVID-19 confirmed cases[1] nor explain the mechanism of superspreading events[30].

Here, we aim to extend the power of the standard SEIR model by introducing a meta-population framework that can consider fine-grained intra-city mobility behavior.

Specifically, we divide the urban space into numerous neighborhoods, and maintain a separate SEIR process for the sub-population in each neighborhood. The overall city population is divided into the neighborhood based on real-world population distributions[27]. In each simulation epoch, our model has two stages to trace epidemic development and population mixing, respectively. In the stage of epidemic development, we compute the changes of population status according to the following equations in each neighborhood:

$$\frac{ds^n}{dt} = -\beta s^n e^n - \beta i^n e^n + \text{Input}_S(t)$$

$$\frac{de^n}{dt} = (1-\omega_i)\beta s^n e^n + (1-\omega_i)\beta s^n i^n - \frac{1}{\tau}e^n + \text{Input}_E(t)$$

$$\frac{di^n}{dt} = \omega_i \beta s^n e^n + \omega_i \beta s^n i^n + \frac{1}{\tau}e^n - \kappa i^n - \gamma i^n + \text{Input}_I(t)$$

$$\frac{dr^n}{dt} = \kappa i^n + \gamma i^n + \text{Input}_R(t)$$

where $s^n, e^n, i^n, r^n$ are the susceptible, exposed, infected and recovered people in neighborhood $n$. $\beta$ is the infection rate, $\omega_i$ is the ratio that secondary infections immediately show symptoms and change to infected status, $\tau$ is average incubation time of exposed persons, and $\kappa$ denotes the quarantine rate that infected persons are removed from the population.

For the infection states for the whole city, we have

$$S = \sum_n s^n$$
$$E = \sum_n e^n$$
$$I = \sum_n i^n$$
$$R = \sum_n r^n$$

Besides, the input terms are determined based on the incoming and outgoing population due to intra-city mobility. Specifically, in the stage of population mixing, we simulate the intra-city mobility behavior based on the following gravity model:

$$m_{ij} = M \frac{N_i^\rho N_j^\theta}{\exp(d_{ij}/r)},$$

where $M$ is the mobility level depicting the intensity of intra-city mobility, $N_i$ and $N_j$ are the population size of the originated and designated neighborhoods, and $d_{ij}$ is the Manhattan distance between them. $\rho, \theta, r$ are empirical coefficients we set based on recent urban mobility research[23]. This gravity model assumes the mobility flows between neighborhoods are negatively correlated travel distance and positively correlated with the population size, which is inspired by Stouffer's law of opportunity-driven movement[24]. We assume the mobility flows have similar distribution of epidemic status as the originated neighborhoods and sum them up as the input terms in designated neighborhoods. We set the infection rate $\beta$, quarantine rate $\kappa$ and mobility level $M$ as learnable parameters to estimate the efficacy of intervention measures like social distancing, mass testing and quarantine, and stay-at-home order. Besides, we set other epidemiological parameters based on the recent COVID-19 research (see Extended Data Table.1).

**M2 Theoretical analysis of complex growth patterns**

To evaluate our model's capacity in capturing the complex growth patterns of COVID-19 confirmed cases, we consider two extreme scenarios of our model. First, we consider a complete lockdown scenario ($M \to 0$) that the coronavirus is contained in the several initial neighborhoods. Therefore, the effective population is the combined population of these neighborhoods, which will be significantly smaller than the whole population ($N_{eff} \ll N$). It is equivalent to an exponential depletion of susceptible people that resembles the self-containment mechanism in the recent SIR-X model, which has been proved that can reproduce sub-exponential growth curves[1]. Moreover, our model provides a micro foundation to explain the sub-exponential growth of COVID-19 cases from the urban mobility behavior perspective.

Another extreme case is the highly efficient city-wide population mixing ($M \to \infty$). In this scenario, the population mixing stage in each simulation epoch will make the persons of different status will distribute proportionately across the neighborhoods based on the size of each subpopulation. Therefore, the epidemic development stage of each neighborhood run a similar SEIR model but with different population sizes:

$$\frac{ds^n}{dt} = -\beta s^n e^n - \beta i^n e^n$$

$$\frac{de^n}{dt} = (1-\omega_i)\beta s^n e^n + (1-\omega_i)\beta s^n i^n - \frac{1}{\tau}e^n$$

$$\frac{di^n}{dt} = \omega_i \beta s^n e^n + \omega_i \beta s^n i^n + \frac{1}{\tau}e^n - \kappa i^n - \gamma i^n$$

$$\frac{dr^n}{dt} = \kappa i^n + \gamma i^n$$

where $s^n, e^n, i^n, r^n$ have similar proportion composition in each neighborhood. For example, the number of susceptible people in neighborhood $n$ can be computed as

$$s^n = \frac{s^n + e^n + i^n + r^n}{S+E+I+R} S$$
$$= p_n S$$

where $p_n$ is defined as the proportion of population in neighborhood $n$.

Therefore, the city-wide epidemic development is a linear sum of a set of homogeneous SEIR models. For example, the coronavirus transmission (the decrease of susceptible population) occur in each simulation epoch can be computed as follow:

$$\frac{dS}{dt} = \frac{ds^1}{dt} + \frac{ds^2}{dt} + \ldots + \frac{ds^m}{dt}$$
$$= -(\beta s^1 e^1 + \beta s^1 i^1) - (\beta s^2 e^2 + \beta s^2 i^2) - \ldots - (\beta s^m e^m + \beta s^m i^m)$$
$$= -(\beta SE + \beta SI)p_1^2 - (\beta SE + \beta SI)p_2^2 - \ldots - (\beta SE + \beta SI)p_m^2$$
$$= -\left(\left(\sum_{i=1}^{m} p_i^2 \beta\right) SE + \left(\sum_{i=1}^{m} p_i^2 \beta\right) SI\right)$$

Similar equations can be derived for exposed, infected, and recovered populations. Therefore, we can see that the city-wide epidemic development is equivalent to a SEIR model with different parameters, where the equivalent infection rate is $\beta_{equ} = \left(\sum_{i=1}^{m} p_i^2 \beta\right)$. As a result, it can reproduce the exponential growth curves in the standard SEIR model.

Our proposed metapopulation model lies between these two extreme cases, with the mobility level parameter ranges from complete lockdown ($M \to 0$) to highly efficient population mixing ($M \to 0$). Therefore, by changing the mobility level, our model can reproduce the complex growth curves ranging from sub-linear to exponential (see Extended Data Fig.3). As a result, our model is provably expressive to capture the complex growth curves of COVID-19 confirmed cases.

## Data availability

All the data that have been used in this work is publicly available. Details of the data are available in the Supplementary Information.

## Acknowledgements

This work was supported in part by The National Key Research and Development Program of China under grant 2020AAA0106000, the National Natural Science Foundation of China under U1936217.

## Author contributions

Fengli Xu, Yong Li, Jianhua Lu and James Evans jointly launch this research. Zhenyu Han performs the experiments and prepares the figures in this work. Fengli Xu, Yong Li, Tao Jiang, Depeng Jin and James Evans provide the research outlines, research designs and critical revisions. All authors jointly analyzed the results and wrote the paper.

## Competing interests

All the authors claim no competing interests.

## Additional information

Supplementary information is available for this paper.

**Extended Data: (Figures and Tables)**

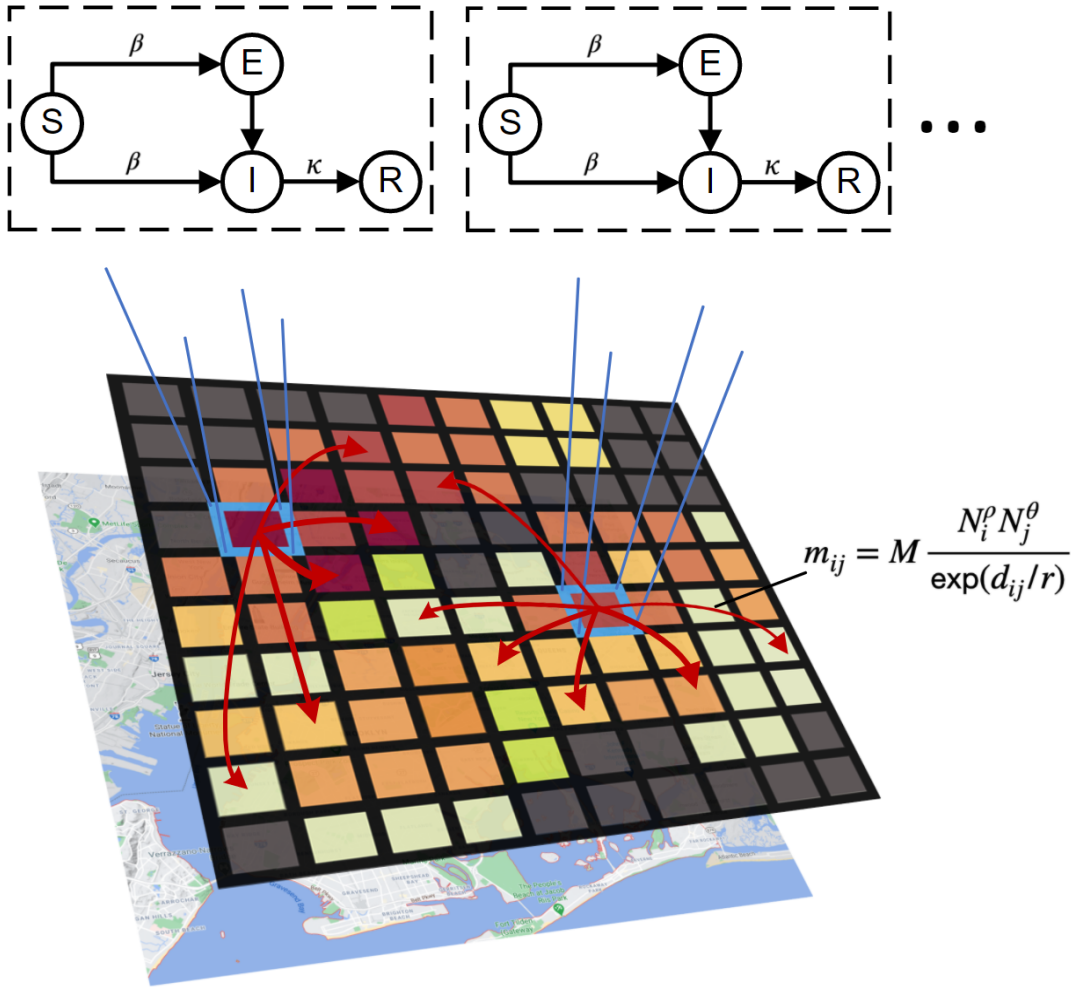

**Extended Data Fig.1| Illustration of the proposed meta-population model.** We segment the urban space into numerous neighborhoods, and divide the whole city population into subpopulation based on them. Separate SEIR model is run on the subpopulation within each neighborhood, and the susceptible, exposed, infected, and recovered individuals are mixed across neighborhoods according to the mobility flow predicted by the gravity model.

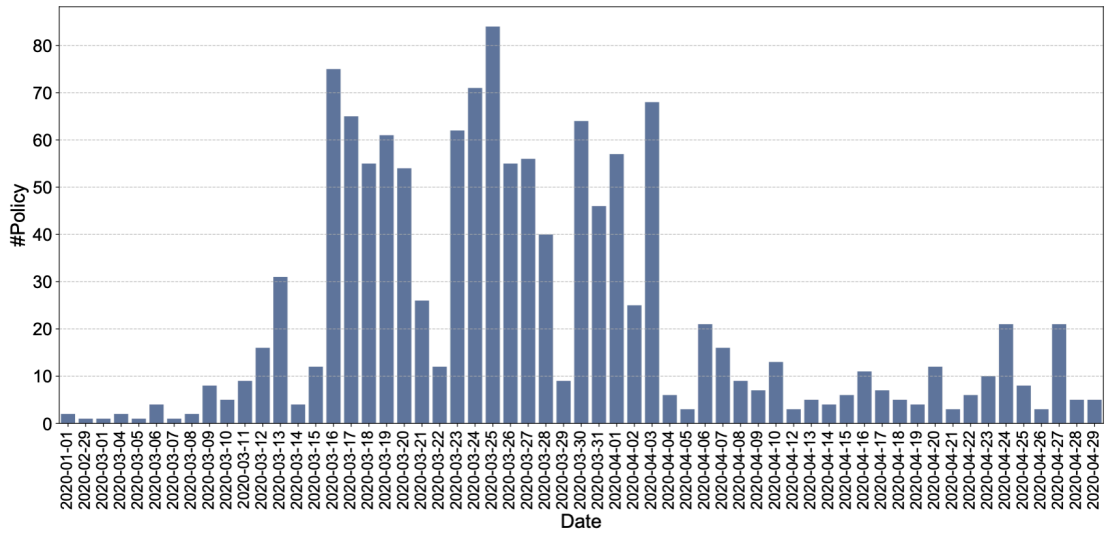

**Extended Data Fig.2| The number of COVID-19 intervention policies announced in the United States.** Most policies are released at the end of March after the Declaration of National Emergency. Thus, we choose April 1st as the time point of behavior change.

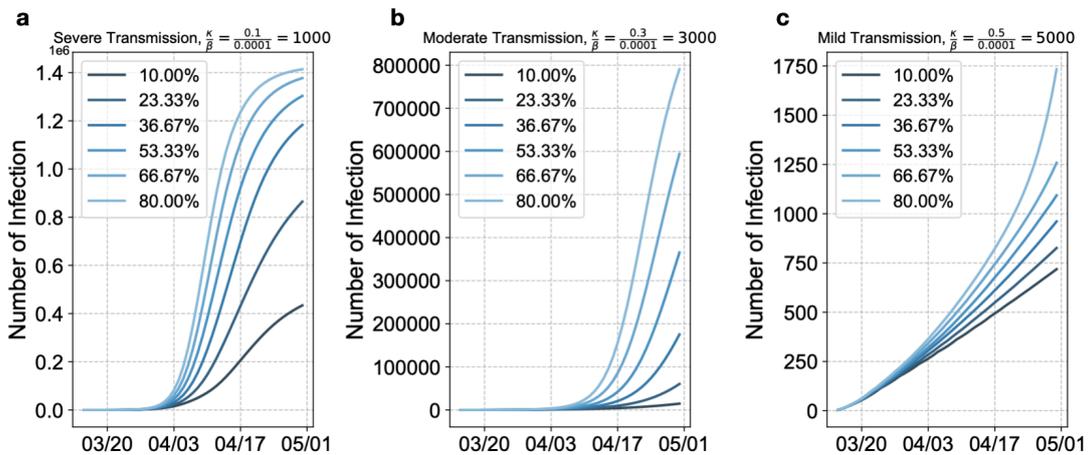

**Extended Data Fig.3| The estimated growth curves of COVID-19 cases with different mobility levels and transmission rates.** Color of lines represent different mobility levels compared with maximum mobility, where all the people are involved in circulation. Different mobility levels will greatly change the shape of growth curves, where low mobility levels are more likely to result in sub-exponential growth curves.

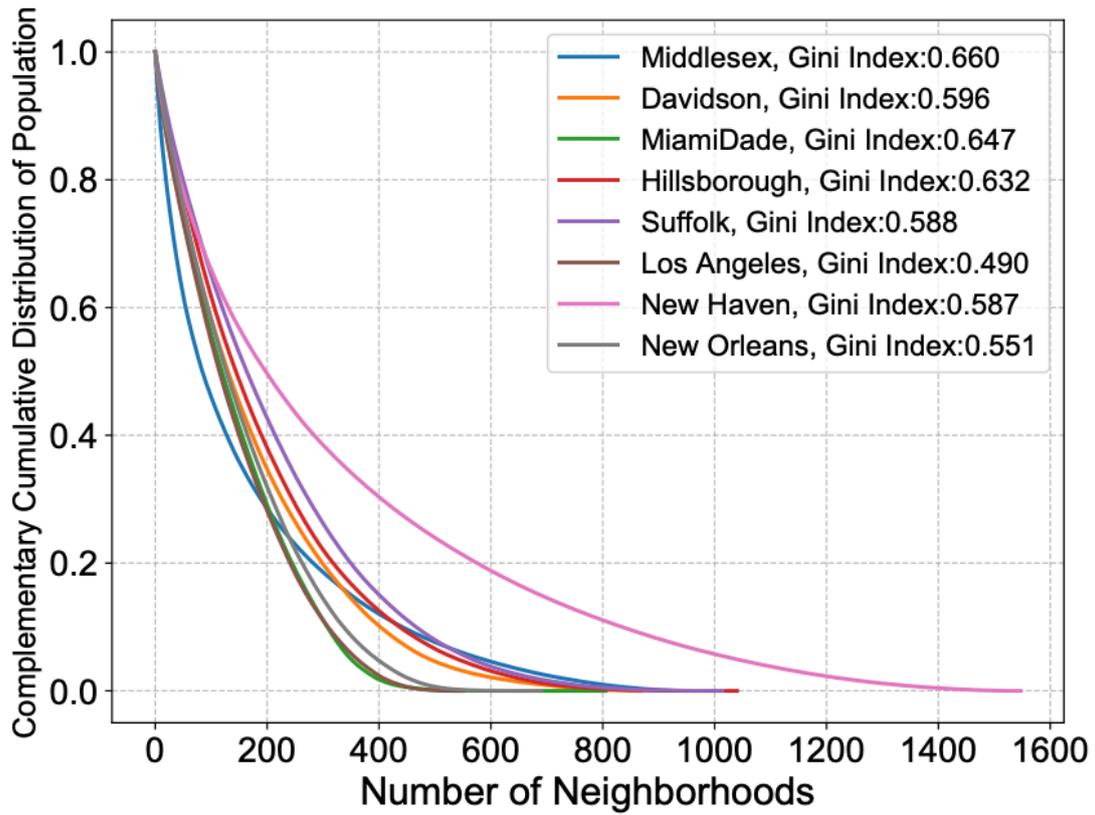

**Extended Data Fig.4| Heterogeneity of population distribution in cities.** Population distributions of selected U.S. counties demonstrate strong heterogeneity where top populated neighborhoods account for considerable population.

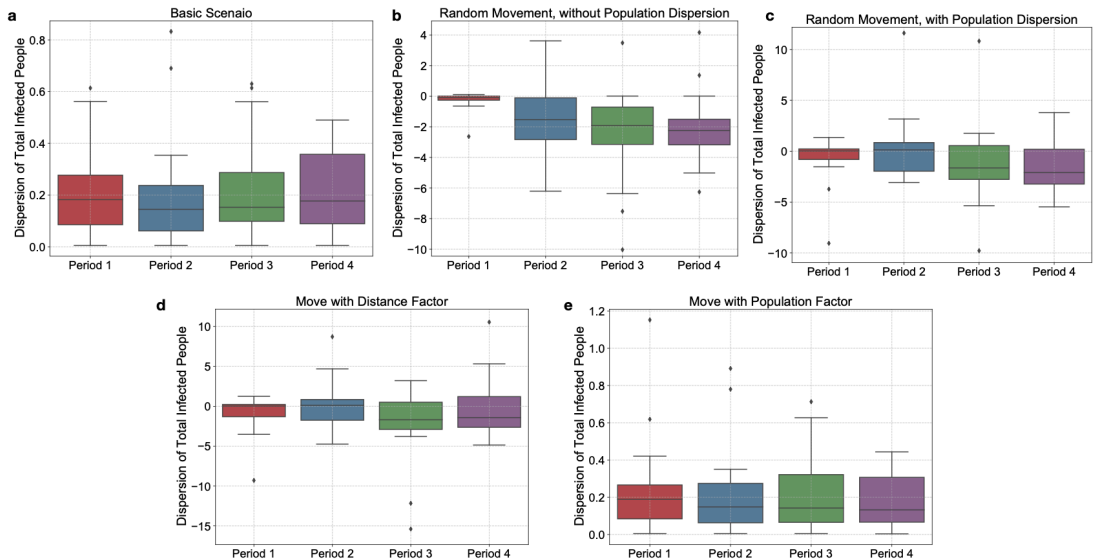

**Extended Data Fig.5| Dispersion parameter of 20 United States counties in different periods of epidemic.** We examine the dispersion parameter of total infected

people (including exposed people and infected people) in different time periods and different scenarios. These periods include the first quarter day, the middle day, the last quarter day and the final day. Smaller dispersion parameter represents greater dispersion. When it is smaller than 1 and greater than 0, significant superspreading exists.

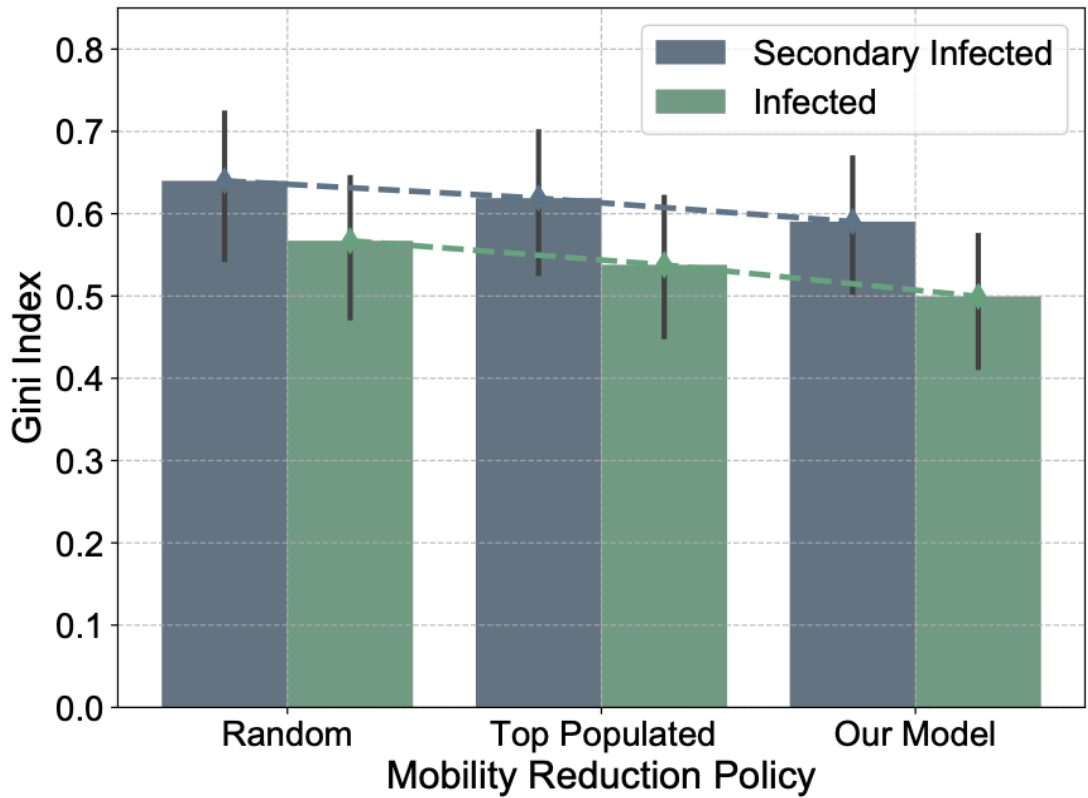

**Extended Data Fig.6 | The effect of superspreading events under different mobility reduction policies.** It is measured by the Gini indexes of the spatial distribution of secondary infections and infected persons. The policy informed by our model results in smaller spatial unevenness of infection risk, which indicates less superspreading events.

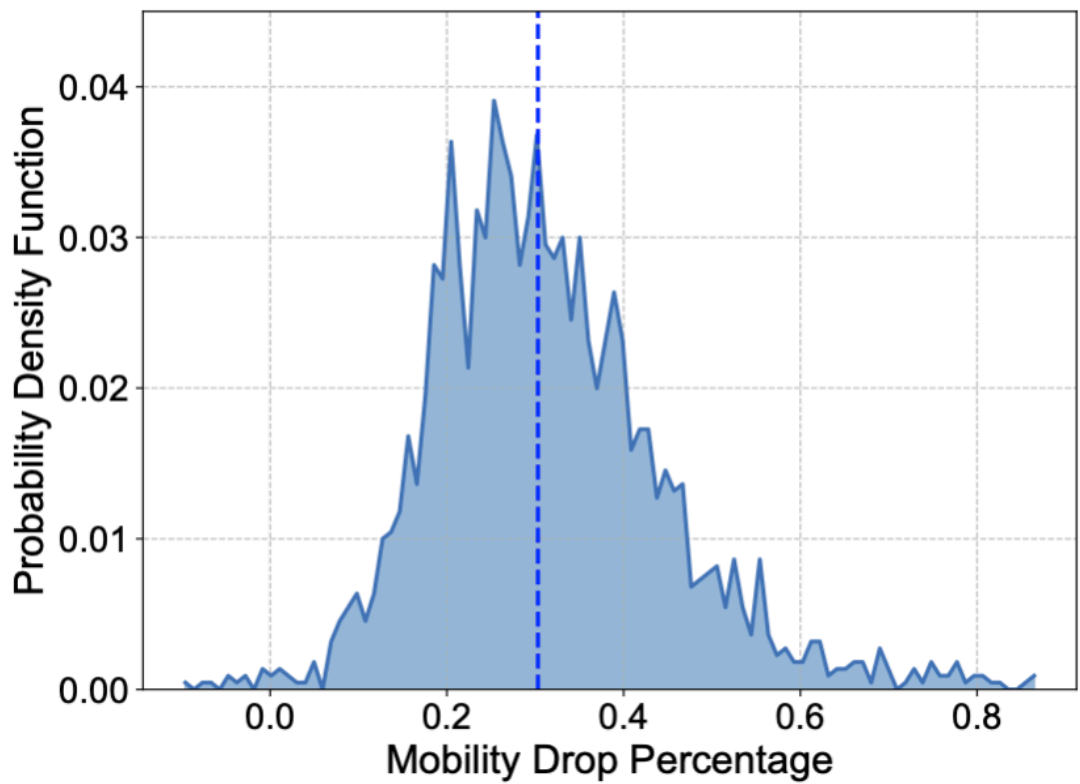

**Extended Data Fig.7 | Mobility drops in selected U.S. counties before and after March 25.** We evaluate the real-world mobility drops from Apple Mobility Trends Reports[10]. The mobility drop has a median value of 30.3% (shown in the vertical dashed line), with IQR 0.227-0.392.

| Name | Description | Value | Source |
|---|---|---|---|
| Infection rate ($\beta$) | Probability of getting infected when a susceptible person collocates with an exposed or infected person. We assume the transmission probability is equal for exposed and infected people. | Learnable parameter | - |
| Quarantine rate ($\kappa$) | The quarantine rate for the infected population. | Learnable parameter | - |
| Mobility level ($M$) | Intra-city mobility level. | Learnable parameter | - |
| Incubation period ($\tau$) | The average incubation period of exposed individuals. | 5.2 | Reference to [3,45,46] |
| Recovery rate ($\gamma$) | The rate of infected individuals recover or die. | 1/14 | According to [47] |
| Probability of secondary infection ($\omega_i$) | The probability that exposed individuals will have a secondary infection, when they directly express the symptom and become infected persons. | 0.7 | Reference to [46] |

**Extended Data Table.1| Epidemiological parameters of our meta-population SEIR model.**

| County | State | Confirmed Cases | Land Area (km²) |
|---|---|---|---|
| New York | New York | 167478 | 783.8 |
| Cook | Illinois | 36513 | 4235 |
| Nassau | New York | 35854 | 1173 |
| Suffolk | New York | 33664 | 6146 |
| Los Angeles | California | 23220 | 1302 |
| Bergen | New Jersey | 15610 | 639 |
| Hudson | New Jersey | 14916 | 162 |
| Philadelphia | Pennsylvania | 14468 | 400 |
| Middlesex | Massachusetts | 14208 | 2195 |
| Union | New Jersey | 12578 | 940 |
| Miami-Dade | Florida | 12063 | 6297 |
| New Haven | Connecticut | 7536 | 2233 |
| Oakland | Michigan | 7267 | 2349 |
| New Orleans | Louisiana | 6452 | 906 |
| Ocean | New Jersey | 6375 | 2370 |
| Harris | Texas | 6356 | 4602 |
| King | Washington | 6207 | 5975 |
| Davidson | Tennessee | 2612 | 1363 |
| Will | Illinois | 2492 | 2200 |
| Hillsborough | Florida | 1124 | 3279 |

**Extended Data Table.2|** The population size and confirmed cases of selected counties until April 30.